\documentclass[12pt]{iopart}

\usepackage{iopams}  
\usepackage{graphicx} 
\usepackage{subfig}
\usepackage{multirow}
\usepackage{dsfont}
\usepackage{cite}
\usepackage{xcolor}

\begin{document}

\title[Recursion scheme for the largest  $\beta$-Wishart-Laguerre eigenvalue]{Recursion scheme for the largest  $\beta$-Wishart-Laguerre eigenvalue and Landauer conductance in quantum transport}

\author{Peter J. Forrester}
\address{School of Mathematics and Statistics, ARC Centre of Excellence for Mathematical and Statistical Frontiers, 
The University of Melbourne,Victoria 3010, Australia}
\ead{pjforr@unimelb.edu.au}

\author{Santosh Kumar}
\address{Department of Physics, Shiv Nadar University, Uttar Pradesh 201314, India}
\ead{skumar.physics@gmail.com}

\begin{abstract}
The largest eigenvalue distribution of the Wishart-Laguerre ensemble, indexed by Dyson parameter $\beta$ and Laguerre parameter $a$, is fundamental in multivariate statistics and finds applications in diverse areas. Based on a generalisation of the Selberg integral, we provide an effective recursion scheme to compute this distribution explicitly in both the original model, and a fixed-trace variant, for $a,\beta$ non-negative integers and finite matrix size. 
For $\beta = 2$ this circumvents known symbolic evaluation based on determinants which become impractical for large dimensions. Our exact results have immediate applications  in the areas of multiple channel communication and bipartite entanglement. Moreover, we are also led to the exact solution of a long standing problem of finding a general result for Landauer conductance distribution in a chaotic mesoscopic cavity with two ideal leads. Thus far, exact closed-form results for this were available only in the Fourier-Laplace space or could be obtained on a case-by-case basis.
\end{abstract}

%
%
%
%
%

\section{Introduction}

The Wishart-Laguerre ensemble constitutes an important class of random matrices and has been of immense usefulness in modeling a variety of problems in diverse topics~\cite{Forrester2010,Mehta2004,Handbook2011}. The corresponding extreme eigenvalues, aside from being fundamental in multivariate statistics~\cite{Constantine1963,Anderson1963,Khatri1964,Muirhead1982,Forrester1993,Forrester1994,FH1994,NF1998,Johansson2000,Johnstone2001,Forrester2007,Forrester2010,WG2013}, play crucial roles in problems ranging from quantum entanglement in bipartite systems~\cite{MBL2008,Majumdar2011,AV2011,KSA2017}, and electronic transport properties in mesoscopic systems~\cite{FH1994,Vivo2011}, to multichannel communication in wireless networks~\cite{KA2003,DMJ2003,MA2005,PL2008,ZCW2009,WT2011,WTDM2012,KSRZ2012,JHCBC2017}. The asymptotic behaviour of these extreme eigenvalues and the corresponding gap probabilities have been explored extensively, leading to Tracy-Widom densities~\cite{TW1993,TW1994,TW1994a,Johansson2000,Johnstone2001,FS2010} and large-deviation results~\cite{VMB2007,MV2009,KC2010,Forrester2012} of significant importance. In certain applications, however, one requires to go beyond universal results and seek exact and explicit computation of these distributions for finite matrix size. Typical theory in this connection is based on determinants or Pfaffians requiring symbolic computation~\cite{Khatri1964,FH1994,NF1998,Forrester2007,ZCW2009,WG2013}, and therefore becomes impractical to evaluate if large matrices are involved. An alternative approach for obtaining explicit expressions involves implementing certain recurrences. This has turned out to be very effective considering the availability of modern software packages which can handle recursive symbolic computation very efficiently. For the smallest eigenvalue of real Wishart-Laguerre matrices, in Refs.~\cite{Edelman1989,Edelman1991} Edelman provided a recursion scheme which has subsequently been generalised to other ensembles and symmetry classes~\cite{FH1994,Forrester1993a,Forrester2010,Forrester2010a,FR2012,KSA2017,FT2019,Kumar2019}. However, extending this to the largest eigenvalue distribution has remained elusive due to its comparatively more convoluted mathematical structure~\cite{Forrester2010}.

In this work, we accomplish the task of providing an efficient recursion scheme for evaluating the probability density function (PDF) as well as the cumulative distribution function (CDF) of the largest eigenvalue of the integer $\beta$-Wishart-Laguerre ensemble for both unrestricted and fixed trace variants. These results apply at once to problems pertaining to multiple channel communication in wireless systems and bipartite entanglement in random pure states. Additionally, we exploit the connection of the largest eigenvalue of the fixed trace Wishart-Laguerre ensemble to the quantum conductance problem and solve a long-standing problem of obtaining a general closed-form result for the Landauer conductance distribution in a chaotic mesoscopic cavity with ideal leads. Thus far, exact closed-form results for this distribution were available only in the Fourier-Laplace space as a determinant or could be obtained on a case by case basis~\cite{JPB1994,MK2004,BM1994,Beenakker1997,SWS2007,KSS2009,KP2010}.

To present our findings, we begin by providing the general results for the functional form of the PDF and CDF of the Wishart-Laguerre largest eigenvalue, along with a brief discussion of their immediate applicability to the multiple channel communication and bipartite entanglement problems. We then briefly describe the machinery behind our proposed recursive approach and also present comparison of analytical and Monte Carlo simulation based results for a few examples. The application to the quantum conductance problem is discussed next, where we point out the implementation of recursion scheme to obtain exact Landauer conductance distribution. The analytical results in this case are contrasted with the numerical results obtained with the aid of scattering matrices modelled using circular ensemble of random matrices. Finally, we conclude with a brief summary of our work.

\section{Exact distribution for the $\beta$-Wishart-Laguerre largest eigenvalue}

The classical cases of Wishart-Laguerre ensemble include real, complex and quaternion positive-definite matrices of the form $GG^\dag$, designated by the Dyson index $\beta=1,2$ and 4, respectively. In recent years, matrix models for continuous $\beta>0$ variants  have been also worked out and have drawn considerable attention. In the general case, the joint PDF for Wishart-Laguerre eigenvalues ($0<x_1,...,x_n<\infty$) is given by 
\begin{equation}
\label{WLPDF}
\mathcal{P}(x_1,\dots,x_n)={1 \over W} \prod_{l=1}^n x_l^a e^{-\beta x_l/2}  \prod_{1 \le j < k \le n} | x_k - x_j|^\beta,
\end{equation}
with $a>-1$. The partition function $W$ is known using the Selberg integral as~\cite{Forrester2010,Mehta2004}
\begin{equation}
W=\left(\frac{2}{\beta}\right)^{\!\gamma}\,\prod_{j=0}^{n-1}\frac{\Gamma\left(\frac{\beta(j+1)}{2}+1\right)\Gamma\left(\frac{\beta j}{2}+a+1\right)}{\Gamma\left(\frac{\beta}{2}+1\right)},
\end{equation}
where $\gamma=n[a+\beta(n-1)/2+1]$.
The PDF and CDF of the largest eigenvalue are computed as~\cite{Forrester2010}
\begin{eqnarray}
\label{PxQx1}
&&P(x)=n\int_0^x dx_2\dots \int_0^x dx_n \mathcal{P}(x,x_2,\dots, x_n),\\
\label{PxQx2}
&&Q(x)=\int_0^x dx_1 \cdots \int_0^x dx_n \, \mathcal{P}(x_1,...,x_n)=\int_0^x dx' P(x').
\end{eqnarray}
The CDF of the largest eigenvalue coincides with the gap probability $ E_{n,\beta}(0;(x,\infty))$ of finding no eigenvalue in the domain $(x,\infty)$.

We demonstrate below that for positive integer $\beta$ and non-negative integer $a$, the sought PDF and CDF exhibit respective structures:
\begin{equation}
\label{Px}
P(x)=\sum_{j=1}^n e^{-j \beta x/2} \sum_{k=a}^{ja+ j(n-j)\beta} c_{jk}x^k,
\end{equation}
\begin{equation}
\label{Qx}
Q(x)=\sum_{j=0}^n e^{-j \beta x/2} \sum_{k=0}^{ja+ j(n-j)\beta} d_{jk}x^k.
\end{equation}
For complex matrices ($\beta=2$), the above structures have been pointed out in Ref.~\cite{DMJ2003} in the context of multiple channel communication problem. However, the computation of the coefficients (which depend on $\beta, a$ and $n$ and evaluate to rational numbers) has remained a daunting task~\cite{DMJ2003,MA2005,WTDM2012}. These earlier works have relied on evaluating the coefficients using the Hankel determinant of an $n$-dimensional matrix. This becomes prohibitive if $n$ becomes large (say, $\gtrsim 10$). As we will see below, our recursive approach facilitates the evaluation of PDF, CDF and coefficients in an efficient manner.
 
The fixed-trace variant of the Wishart-Laguerre ensemble is described by the joint density
 \begin{equation}
 \label{Pft}
\mathcal{P}_F(y_1,\dots, y_n)={1 \over W_F}\delta\left(\sum_i y_i-1\right) \prod_{l=1}^n y_l^a \prod_{1 \le j < k \le n} | y_k - y_j|^\beta,
\end{equation}
where the eigenvalues $(0\leq y_1,...,y_n\leq 1)$ are constrained by the Dirac-delta condition, and $W_F=[1/\Gamma(\gamma)](\beta/2)^\gamma W$ is the corresponding partition function. The PDF and CDF of the largest eigenvalue for this ensemble follows from the Fourier-Laplace relationship with the unrestricted trace variant~\cite{Forrester2010}, equation~(\ref{WLPDF}), as
\begin{equation}
 \label{Pfy}
 \fl
 ~~~~~~~~~P_F(y)=\frac{2}{\beta}\Gamma(\gamma)\sum_{j=1}^n \Theta(1-j y) \sum_{k=a}^{ja+ j(n-j)\beta} \frac{c_{jk}}{\Gamma(\gamma-k-1)}\left(\frac{2y}{\beta}\right)^k (1-jy)^{\gamma-k-2},
\end{equation}
\begin{equation}
\label{Qfy}
Q_F(y)=\Gamma(\gamma)\sum_{j=0}^n \Theta(1-j y) \sum_{k=0}^{ja+ j(n-j)\beta} \frac{d_{jk}}{\Gamma(\gamma-k)}\left(\frac{2y}{\beta}\right)^k (1-jy)^{\gamma-k-1},
\end{equation}
where $\Theta(z)$ denotes the Heaviside step function.
 It should be noted that the largest eigenvalue in this case also coincides with the scaled largest eigenvalue associated with~(\ref{WLPDF}), i.e., $y_\mathrm{max}=x_\mathrm{max}/(x_1+\dots+x_n)$.
 
The above results apply directly to the multiple channel communication and bipartite entanglement problems.
 In multiple input multiple output (MIMO) communication, the channel matrix $H$ is the central object. It models the fading of the signal between $n_t$ and $n_r$ number of transmitting and receiving antennas, respectively. The singular values of the $n_r\times n_t$--dimensional $H$ matrix or, equivalently, the eigenvalues of $HH^\dag$ (or $H^\dag H)$ play a crucial role in deciding several quantities of interest which assess the performance of the MIMO system. In the cases of one-sided Gaussian and Rayleigh fadings, the eigenvalues are described by the density given in~(\ref{WLPDF}) with $\beta=1$ and 2, respectively, and the parameter $a=\beta(|n_t-n_r|+1)/2-1$.  In Ref.~\cite{DMJ2003, MA2005}, it has been shown within a transmit-beamforcing model with maximum ratio combining (MRC) receivers that the maximum output signal-to-noise (SNR) is directly related to the largest eigenvalues of  $HH^\dag$. Consequently, the distribution of the largest eigenvalue has been used to obtain the statistics of symbol error probability (SEP), outage probability and ergodic channel capacity~\cite{DMJ2003, MA2005}. Additionally, the distribution of the scaled largest eigenvalue finds applications in various hypothesis testing problems, both in statistics and in signal processing~\cite{JG1972,Davis1972,SKC1973,KS1974,BS2006,Nadler2011,BDMN2011,WTDM2012}. 
  
In the problem of bipartite entanglement, one is interested in the properties of the reduced density matrix. To elaborate, consider Hilbert spaces $\mathcal{H}_n, \mathcal{H}_m$  of dimensions $n$ and $m$ for the constituents with $n\le m$. Moreover, let $|\phi\rangle$ be a pure state belonging to the composite system, i.e., $|\phi\rangle\in \mathcal{H}_n\otimes\mathcal{H}_m$. In case $|\phi\rangle$ is chosen randomly in a uniform manner, the Schmidt eigenvalues of the reduced density matrix $\rho=\tr_m(|\phi\rangle\langle\phi|)$ obtained by partial tracing over the part belonging to $\mathcal{H}_m$ are governed by the joint PDF~(\ref{Pft}) with $a=\beta(m-n+1)/2-1$~\cite{Forrester2010,ZS2001}. The largest of these eigenvalues varies from $1/n$ to 1. In these two extremes, owing to the fixed trace restriction, all other eigenvalues assume a common value of $1/n$ and 0, respectively. These respective scenarios correspond to the two subsystems being maximally entangled and separable. 
The distribution of the largest eigenvalue of the fixed trace ensemble therefore carries important information regarding the entanglement between the subsystems of the composite bipartite system~\cite{Majumdar2011}.

\section{Recursion scheme}

We now briefly describe the machinery behind the recursion scheme to obtain the CDF and PDF of the largest eigenvalue. 
To begin with, we will demonstrate that the multi-dimensional integral in~(\ref{PxQx2}) for $Q(x)$ does lead to the structure shown in~(\ref{Qx}). For even $\beta$, we expand $ \prod_{j < k} | x_k - x_j|^\beta$ as a homogeneous multivariable
polynomial of degree $\beta n (n-1)/2$, viz.,
\begin{equation}
\label{vandexp}
 \prod_{1 \le j < k \le n} | x_k - x_j|^\beta = \sum_{\kappa = (\kappa_1,\dots, \kappa_n)} \alpha_\kappa \prod_{l=1}^n x_l^{\kappa_l},
\end{equation}
for some coefficients $\alpha_\kappa$.
Here $\kappa$ is an integer partition involving up to $n$ parts and of fixed length such that $\sum_{i=1}^n  \kappa_i = \beta n (n-1)/2$. With $\kappa_1 \ge \kappa_2 \ge \cdots $, it is known $\kappa_j \le (n - j)\beta$. Substituting the above expansion in~(\ref{PxQx2}) yields
\begin{equation}
\label{11}
Q(x)=\frac{1}{W} \sum_\kappa \alpha_\kappa \prod_{l=1}^n \left( \int_0^x dx_l \, x_l^{a+\kappa_l} e^{-\beta x_l/2}\right).
\end{equation}
Provided $a$ is a non-negative integer, the integral inside the above product can be evaluated as
\begin{equation}
\label{intgrl}
 \int_0^x ds\,s^a e^{-\beta s/2} = a! \left(\frac{2}{\beta}\right)^{a+1}\left[ 1 - e^{-\beta x/2} \sum_{k=0}^a {1\over k!} \left(\frac{\beta x}{2}\right)^k\right].
 \end{equation}
Therefore, eventually we wind up with the expansion of the form given in~(\ref{Qx}), but with the upper limit of the inner sum (over $k$) equal to $ja+j[n-(j+1)/2]\beta$. As shown in the Appendix A, this can be refined to $ja + j(n - j)\beta$ as in~(\ref{Qx}). Also, Eqs.~(\ref{11}) and (\ref{intgrl}) imply that for $\beta$ even $d_{j0} = (-1)^j \left({n\atop j}\right)$.
 For handling odd $\beta$, we order the eigenvalues so that  $x_1>x_2>\dots x_n>0$, and then the expansion in~(\ref{vandexp}) is valid for appropriate coefficients $\alpha_\kappa$. Moreover, the $Q(x)$ can be written as the iterated multi-dimensional integral $n! \int_0^x dx_1  \cdots \int_0^{x_{n-2}} dx_{n-1}\int_0^{x_{n-1}} dx_n \mathcal{P}(x_1,...,x_n)$,
leading again to~(\ref{Qx}). The expansion in~(\ref{Px}) for the PDF $P(x)$ follows similarly by using Eqs.~(\ref{vandexp}) and~(\ref{intgrl}) in~(\ref{PxQx1}).

We next discuss a recursion scheme for the computation of the coefficients.
We introduce the auxiliary function, generalising the Laguerre weighted Selberg
integral~\cite{Forrester2010,Mehta2004}
\begin{eqnarray}
\nonumber
	&&	L_{p,\nu}^{(\alpha)}(x) = {p! (\nu-p)! \over \nu! }  \int_{0}^x dt_1 \cdots \int_{0}^x dt_\nu \,\prod_{l=1}^\nu t_l^{\lambda_1} e^{-\lambda t_l} |x-t_l|^{\alpha}
		\\
	&&	\times  \prod_{1\leq j<k \leq \nu} | t_k -t_j|^{2\lambda} e_p(x-t_1,\ldots,x-t_\nu),
\end{eqnarray}
	where $e_p(t_1,\ldots,t_\nu)$ are the elementary symmetric polynomials. At the heart of the recursion lies a remarkable linear differential-difference equation \cite{FR2012,FT2019},
\begin{eqnarray}	
\label{recur}
\nonumber
	&&\lambda (\nu-p) L_{p+1,\nu}^{(\alpha)}(x) = [\lambda(\nu-p)x+B_p] L_{p,\nu}^{(\alpha)}(x) \\
	&&+ x{d\over dx}L_{p,\nu}^{(\alpha)}(x) - D_p x L_{p-1,\nu}^{(\alpha)}(x),
\end{eqnarray}
where $p=0,1,...,\nu-1$ and
	\begin{eqnarray*}
		B_p &= (p-\nu)[\lambda_1 + \alpha +1+ \lambda(\nu-p-1)],~~~
		D_p &= p[\lambda(\nu-p)+\alpha+1].
	\end{eqnarray*}
Suppose $L_{0,\nu}^{(\alpha)}(x)$ is known. Application of the recurrence allows for the computation of  $L_{\nu,\nu}^{(\alpha)} (x)$ which is identical to $L_{0,\nu}^{(\alpha+1)} (x)$. 
With $\nu = n$, $2 \lambda = \beta$, iterating to the value $\alpha = \beta$ gives $Q(x)$.
	
\begin{figure}[!t]
\centering
\includegraphics[width=0.8\linewidth]{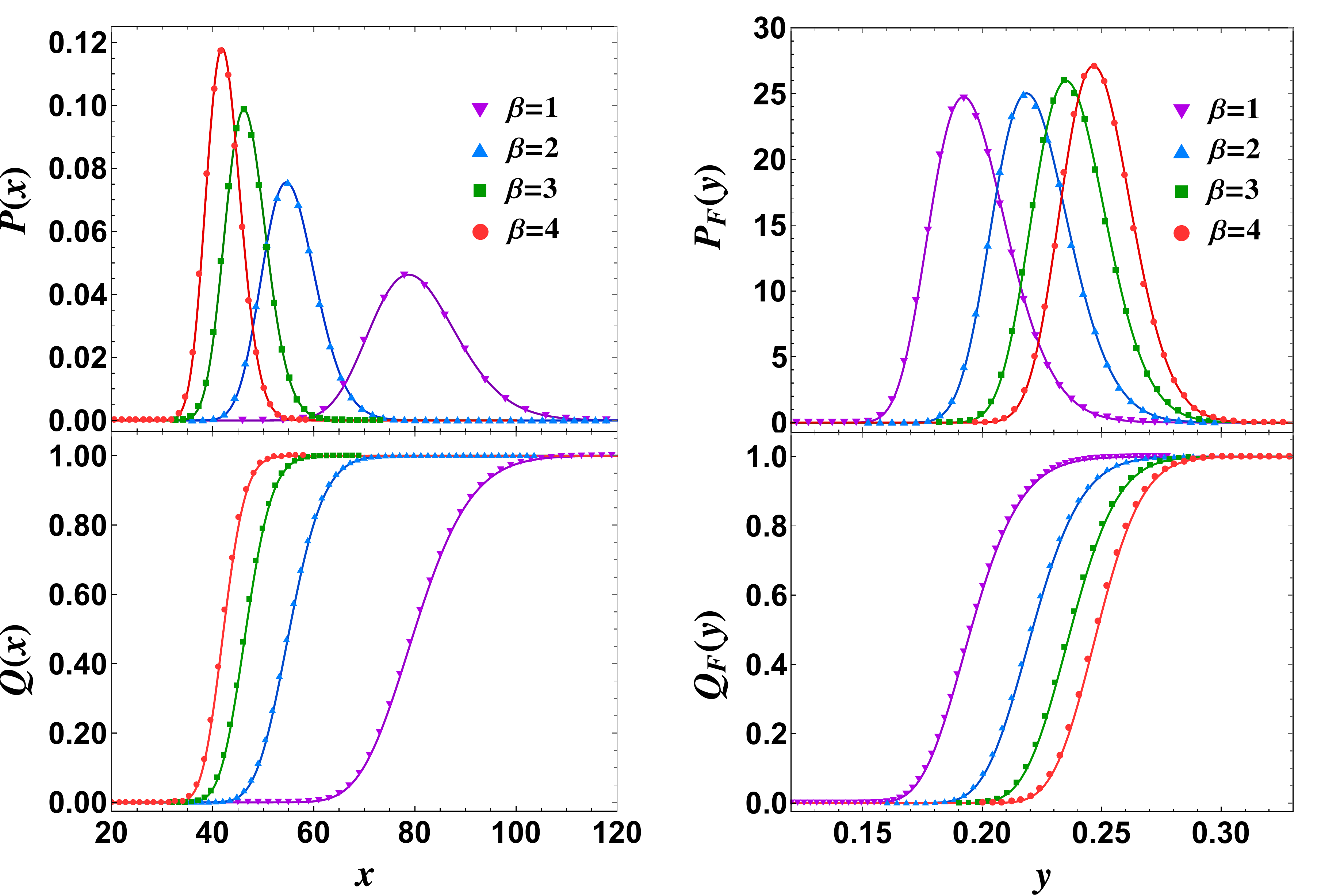}
\caption{PDF and CDF of the largest eigenvalue for the unrestricted trace (left) and the fixed trace (right) Wishart-Laguerre ensembles for $n=10, a=15$.}
\label{utwl}
\end{figure}	
	
The difficulty is that $L_{0,\nu}^{(\alpha)}(x)$ is known only for $\nu = 1$. Fortunately, the relation between $P(x)$ and $Q(x)$ displayed in~(\ref{PxQx2}) permits an iteration also in $\nu$, as observed in
a recursive computation of the largest eigenvalue PDF of the Gaussian orthogonal ensemble given by James~\cite{James1975}. Thus  we begin with~(\ref{intgrl}) to obtain $\int_0^x dx_1\, x_1^a e^{-\beta x_1/2}$. The recurrence in~(\ref{recur}) can then be applied $\beta$ times to evaluate
	$$
	\int_0^x  dx_1\,x_1^a e^{-\beta x_1/2}  |x - x_1|^\beta.
	$$
	Next, we denote $x$ by $x_2$ in the above expression, multiply by $x_2^a e^{-\beta x_2/2}$, and integrate over $x_2$ from 0 to $x$ to obtain
	$$
	 \int_0^{x} dx \, x_2^a e^{-\beta x_2/2}   \int_0^{x_2} dx_1 \, x_1^a e^{-\beta x_1/2}  |x_2 - x_1|^\beta,
	$$
	which is the same as half this double-integral with the upper limit of the inner integral changed to $x$ from $x_2$.
	Now we essentially repeat this procedure. Using the recurrence, knowledge of the above two dimensional
	integral is used to compute
	$$
	 \int_0^{x} \!\!dx_2 \, x_2^a e^{-\beta x_2/2}   \int_0^{x} \!\! dx_1 \, x_1^a e^{-\beta x_1/2}  \prod_{1\le j<k\le3}|x_k- x_j|^\beta
	 $$
	 with $x_3=x$, which is then multiplied by $x_3^a e^{-\beta x_3/2}$, and integrated over $x_3$,
and so on. Continuing, we eventually arrive at the expansion~(\ref{Qx}). An important point is that at all stages of the iteration the one-dimensional integral in~(\ref{intgrl}) applies and the expressions can be written in an expansion analogous
	to~(\ref{Qx}). It should be also noted that the expansion for $P(x)$ appears in this procedure a step before the final integration is considered for $Q(x)$. Furthermore, when evaluating the expressions for a given $n$, the results for lower dimensions are also obtained in the process.
	
A Mathematica~\cite{Mathematica} code to implement the above described recursion scheme can be found in the supplementary material. It also extracts the coefficients $c_{jk}$ and $d_{jk}$ which can then be used for evaluating results for the fixed trace ensemble. As an example, we have considered $n=10, a=15$ and shown the plots for $\beta=1,2,3,4$ in Fig.~\ref{utwl}. The analytical result based solid curves are contrasted against the numerical simulation based overlaid symbols and we can see an excellent agreement.

\section{Exact distribution of Landauer conductance}
One of the remarkable achievements of random matrix theory (RMT) has been in the field of quantum conductance in chaotic mesoscopic systems~\cite{Imry1986,AS1986,SMMP1991,BM1994,JPB1994,Beenakker1997,MK2004,Beenakker2015}. Starting from its prediction of universal conductance fluctuation~\cite{Imry1986,AS1986,SMMP1991,BM1994,Beenakker1997} to the modern day description of topological superconductors and Majorana fermions~\cite{Beenakker2015}, RMT has had great success in modeling various charge transport related phenomena. Despite this, there remain several problems which have defied an exact solution. One such problem is working out the exact distribution of Landauer conductance in a chaotic mesoscopic cavity with ideal leads~\cite{JPB1994,MK2004,BM1994,Beenakker1997,SWS2007,KSS2009,VMB2008,KP2010}. Here we solve this problem by exploiting its mathematical connection with largest eigenvalue of the fixed trace Wishart-Laguerre ensemble.

The chaotic mesoscopic cavity is connected to an electron reservoir via two ideal leads. The electronic transport properties follow from the knowledge of the scattering matrix ($S$ matrix). In this case, the $S$ matrix can be modelled using Dyson's circular ensemble~\cite{BM1994,JPB1994,Beenakker1997,Forrester2006} (note that for nonideal leads the $S$-matrix has to be taken from a nonuniform measure given by the Poisson kernel~\cite{Brouwer1995,Beenakker1997,Forrester2010,VK2012}). For given number of channels $n_1$ and $n_2$ in the two leads, the $S$ matrix is $n_1+n_2$ dimensional. The Landauer conductance, measured in units of $e^2/h$, is then given by $g=\mathrm{tr}(\Pi_1 S\Pi_2 S^\dag)$, where $\Pi_1$ is a projection matrix with elements $(\Pi_1)_{j,k}=\delta_{jk}$ for $j\le n_1$ and zero otherwise and $\Pi_2=\mathds{1}_{n_1+n_2}-\Pi_1$. For $\beta=4$, the $S$ matrix has quaternionic entries and $\Pi_1,\Pi_2$ are accordingly modified. The Landauer conductance is equivalently described in terms of the transmission eigenvalues $(0\leq T_1,...,T_n\le 1)$ as~\cite{Beenakker1997,Landauer1957,FL1981}
\begin{equation}
g=\sum_{j=1}^n T_j,
\end{equation}
where $n=\min(n_1,n_2)$. We also define $m=\max(n_1,n_2)$. The joint PDF describing the transmission eigenvalues is given by~\cite{BM1994,JPB1994,Beenakker1997,Forrester2006,KP2010a}
\begin{equation}
P(T_1,....,T_n)\propto \prod_{l=1}^n T_l^a  \prod_{1 \le j < k \le n} | T_k - T_j|^\beta,
\end{equation}
where $a=\beta(m-n+1)/2-1$. The above is a special case of the Jacobi ensemble of random matrices~\cite{Forrester2006,Forrester2010,KP2010a}. It should be noted that the exponent $a$ is an integer for $\beta=2,4$ for any number of channels $n_1,n_2$, whereas for $\beta=1$ we require $|n_1-n_2|$ to be an odd integer.  The distribution of the Landauer conductance follows as
\begin{equation}
P_g(g)=\int_0^1 dT_1\cdots\int_0^1 dT_n \delta\left(g-\sum_{j=1}^n T_j\right)P(T_1,\dots,T_n).
\end{equation}
A simple scaling $y_j=T_j/g$ reveals that the above can be written in terms of the CDF of the largest eigenvalue of the fixed trace Wishart-Laguerre ensemble~\cite{Vivo2011},
\begin{equation}
P_g(g)=\frac{1}{\Gamma(\gamma)}K g^{\gamma-1}Q_F(1/g);
\end{equation}
\begin{equation}
K=\prod_{l=0}^{n-1}\frac{\Gamma\Big(a+\frac{\beta(n+l-1)}{2}+2\Big)}{\Gamma(\frac{\beta l}{2}+1)}.
\end{equation}
Plugging in the expression for $Q_F$ from~(\ref{Qfy}) gives
\begin{equation}
\label{Pg}
P_g(g)=K \sum_{j=0}^n \Theta(g-j)
 \sum_{k=0}^{ja+ j(n-j)\beta} \frac{d_{jk}}{\Gamma(\gamma-k)}\left(\frac{2}{\beta}\right)^k(g-j)^{\gamma-k-1}.
\end{equation}
This equation provides the sought exact Landauer conductance distribution for general non-negative integer $a$, $\beta$. Corresponding Mathematica code is provided in the supplementary material. In Fig.~\ref{lcd}, as examples, we show the distribution of scaled conductance, viz., $n P_g(ng)$ vs.~$n g$ for $\beta=1,2,4$ and various $n,m$ combinations. The numerical results obtained using simulation of scattering matrices from circular ensembles are overlaid as symbols on the solid curves based on analytical results. We can see a perfect agreement.
\begin{figure*}[!t]
\centering
\includegraphics[width=\textwidth]{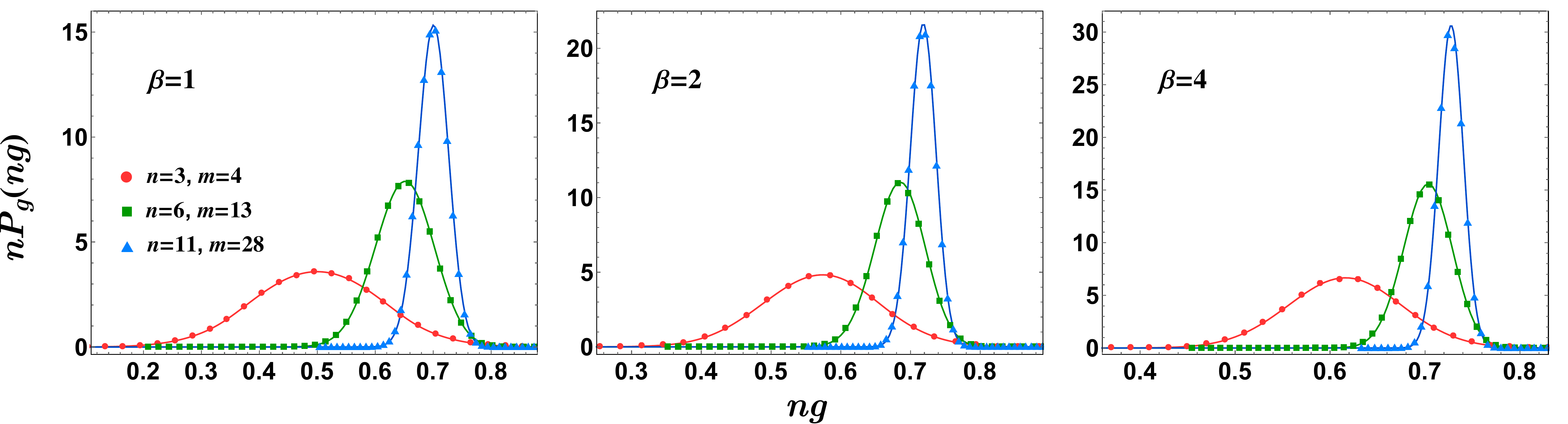}
\caption{Distribution of the Landauer conductance for $\beta=1,2,4$ and three combinations of the number of channels $n_1, n_2$, as indicated using $n=\min(n_1,n_2)$ and $m=\max(n_1,n_2)$.}
\label{lcd}
\end{figure*}

\section{Summary and conclusion}
In this work, we have provided an efficient recursive scheme for the exact  computation of the largest eigenvalue distribution of the integer $\beta$-Wishart-Laguerre ensemble. Applications to multiple channel communication, bipartite entanglement problems, and  an exact solution to the Landauer conductance distribution in a chaotic mesoscopic cavity with ideal leads have been highlighted.

We would like to conclude by emphasising that the recursive approach proposed in this work for the Wishart-Laguerre largest eigenvalue can be obtained from the more general case of the Jacobi ensemble \cite{FR2012,FT2019}. 
The details of this, and its consequences, are planned for a future work.

\section*{Acknowledgements}
P.J.F.~acknowledges support from the Australian Research Council (ARC) through the ARC Centre of Excellence for Mathematical and Statistical frontiers (ACEMS) and ARC grant DP170102028. S.K.~acknowledges the support by the grant EMR/2016/000823 provided by SERB, DST, Government of India. Comments by Mario Kieburg on an earlier draft of the manuscript are appreciated.


\appendix

\section{Refinement of upper limit of inner summation in equation (6)}

We discuss here the refinement of the upper limit of inner summation in~(6). For this we consider the $j$-point correlation function,
 \begin{equation}
 \rho_{(j)}(x_1,\dots, x_j)={n!\over(n-j)!}\int_0^\infty dx_{j+1}\cdots\int_0^\infty dx_n  \mathcal{P}(x_1,...,x_n),
 \end{equation}
and the following expansion of $Q(x)$ in terms of $\rho_{(j)}$:
 \begin{equation}
Q(x) = 1 + \sum_{j=1}^n \frac{(-1)^j}{j!} \int_x^\infty dx_1 \cdots  \int_{x}^\infty dx_j \, \rho_{(j)}(x_1,\dots, x_j).
 \end{equation}
We then use the fact that for large $x_1,\dots, x_j$, $ \rho_{(j)}(x_1,\dots, x_j)$ is proportional to
 \begin{equation}
\prod_{l=1}^j x_l^{a+ \beta (n-j)} e^{-\beta x_l/2}\prod_{1 \le p < q \le j} |x_p - x_q|^\beta.
 \end{equation}
and perform integration by parts multiple times. The latter procedure reveals, for each $j$, the exponent $\kappa$ in the leading
large  $x$ form $x^\kappa e^{- \beta j x/2}$
and thus the upper limit in the summation over $k$ in~(6). The value $\kappa = ja+ j(n-j)\beta$ is therefore obtained, as appears in~(6).


\newpage
\begin{center}
{\bf \large Recursion scheme for the largest $\beta$-Wishart-Laguerre eigenvalue and Landauer conductance in quantum transport: Supplementary Material}
\end{center}
\begin{center}
{\it Peter J. Forrester and Santosh Kumar}
\end{center}

\section*{Description of the Mathematica code}

We implement the proposed recursion scheme in a Mathematica code presented below. 

From the discussion of the recursion scheme, it follows that we need a three level nested loop to obtain the expressions for the PDF $P(x)$ and CDF $Q(x)$. Since the overall expression for a given dimension $n$ builds upon the lower values of dimension, the outermost loop runs over $\nu=1,2,..,n-1$. At a given $\nu$ value, the innermost loop involves recursion over $p$ as in Eq.~(14) from $p=0$ to $\nu-1$, so as to increase the exponent $\alpha$ by 1. The $\alpha$, overall, needs to be iterated from $0$ to $\beta-1$ so as to obtain the result for a given $\beta$. This is achieved in the middle loop. Once these three loops are completed, the expression for $P(x)$ is obtained. A final integration is then performed (corresponding to $\nu=n$) to yield the corresponding $Q(x)$. 

The algorithm involves performing integrals over expressions containing linear combination of terms like $x^r e^{-sx}$ (Laguerre weight) several times. We found that the direct use of the inbuilt function `Integrate' in Mathematica is extremely inefficient at performing this task if there are many terms in the expression. Therefore, we use the `Map (/@)'  function and apply the result in Eq.~(12) to individual terms in the expression and thereby obtain the overall integrated result in view of the linearity. 

The coefficients $c_{jk}$ and $d_{jk}$ are extracted readily from the obtained $P(x)$ and $Q(x)$ expressions with the aid of the `CoefficientList' function in Mathematica. These are then used to yield explicit expressions for the PDF $P_F(y)$ and CDF $Q_F(y)$ for the fixed trace ensemble and also the Landauer conductance density function $P_g(g)$. It should be noted that in Mathematica list indices start from 1 and not from 0, therefore the indices in the Eqs.~(5),~(6),~(8),~(9) and~(20) are accordingly shifted when implemented in the code.

From Eqs.~(5) and~(6) it is clear that there are $(n/6)[(n^2-1)\beta+3(n-1)a+6]$ and $(1/6)(n+1)[n^2\beta+n(3a-\beta)+6]$ terms in the expanded expressions of $P(x)$ and $Q(x)$, respectively. As a consequence, the symbolic expressions become very lengthy even for moderate values of $n$ and $a$. For example, if we consider $\beta=2,n=5,a=5$, the number of terms in the expanded expressions comes out to 95 and 121, respectively. For numerical evaluation of such lengthy expressions we must work with a very high precision so as to prevent under- or over-flow.

The above described three level nested loop algorithm may be contrasted with that for the smallest eigenvalue~[43], which involves a two level nested loop only due to a less involved mathematical structure there.

\begin{figure*}[!t]
\centering
\includegraphics[width=0.95\textwidth]{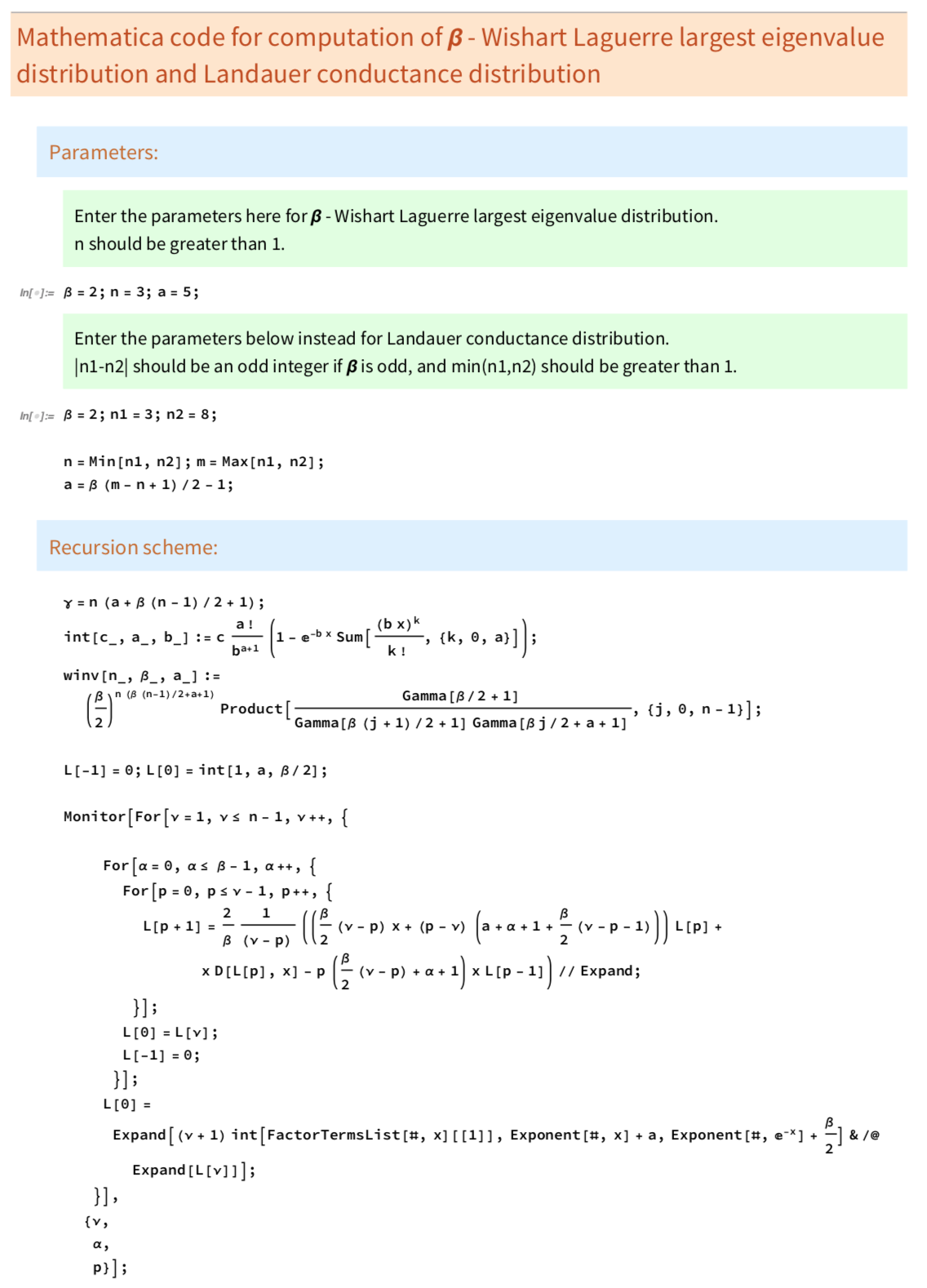}
\end{figure*}
\begin{figure*}[!t]
\centering
\includegraphics[width=0.95\textwidth]{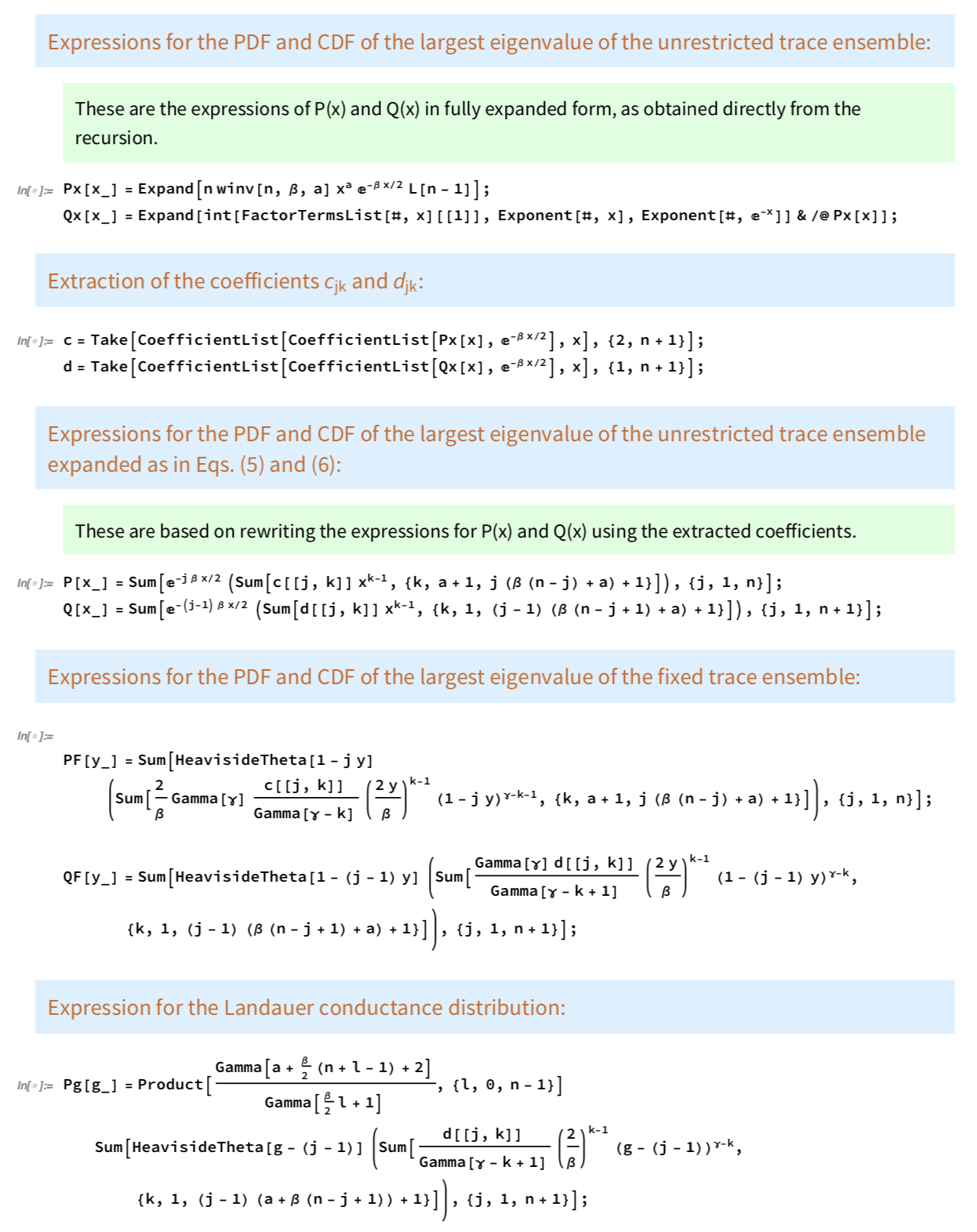}
\end{figure*}
\begin{figure*}[!t]
\centering
\includegraphics[width=0.95\textwidth]{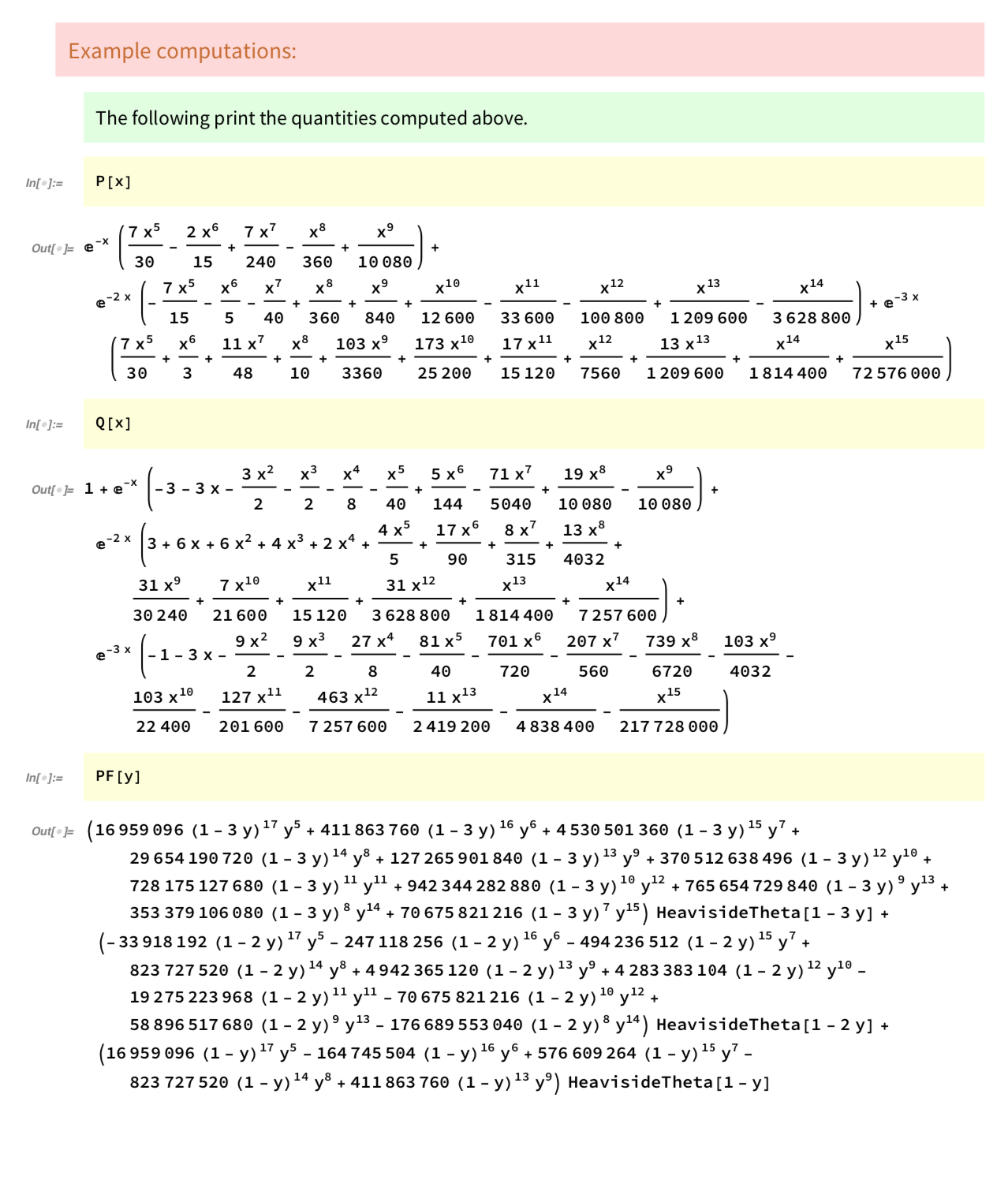}
\end{figure*}
\begin{figure*}[!t]
\centering
\includegraphics[width=0.95\textwidth]{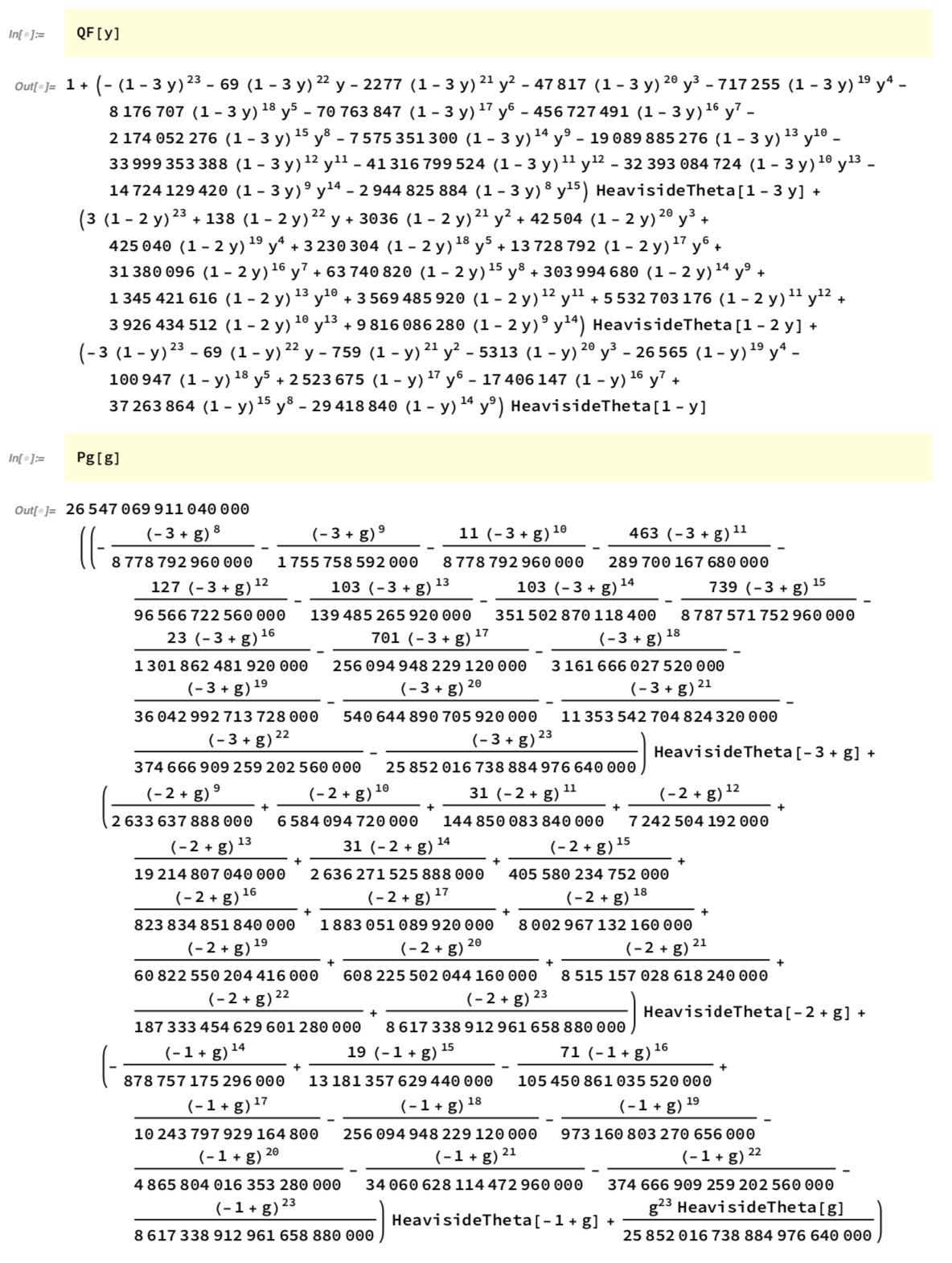}
\end{figure*}
\begin{figure*}[!t]
\centering
\includegraphics[width=0.95\textwidth]{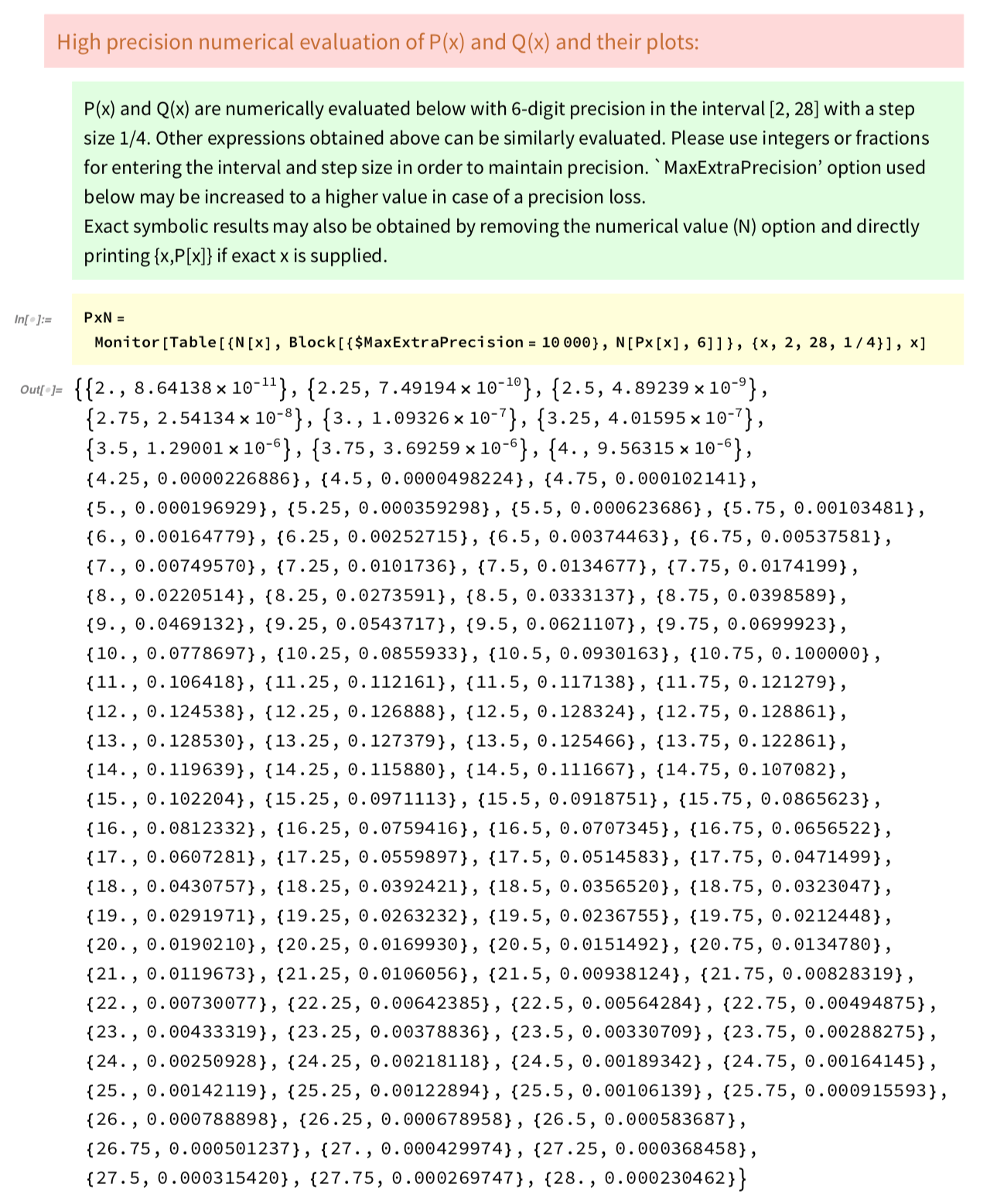}
\end{figure*}
\begin{figure*}[!t]
\centering
\includegraphics[width=0.95\textwidth]{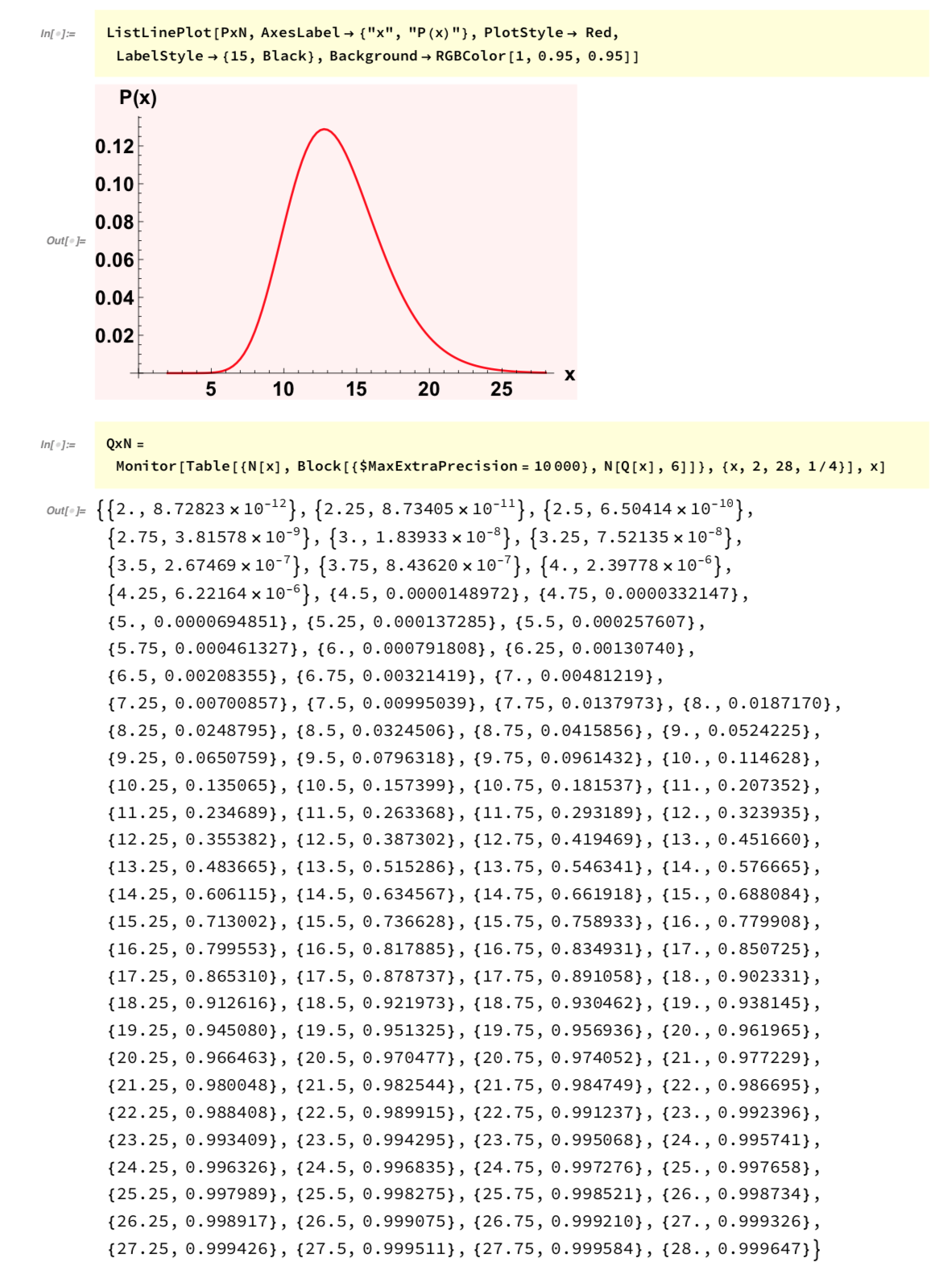}
\end{figure*}
\begin{figure*}[!t]
\centering
\includegraphics[width=0.95\textwidth]{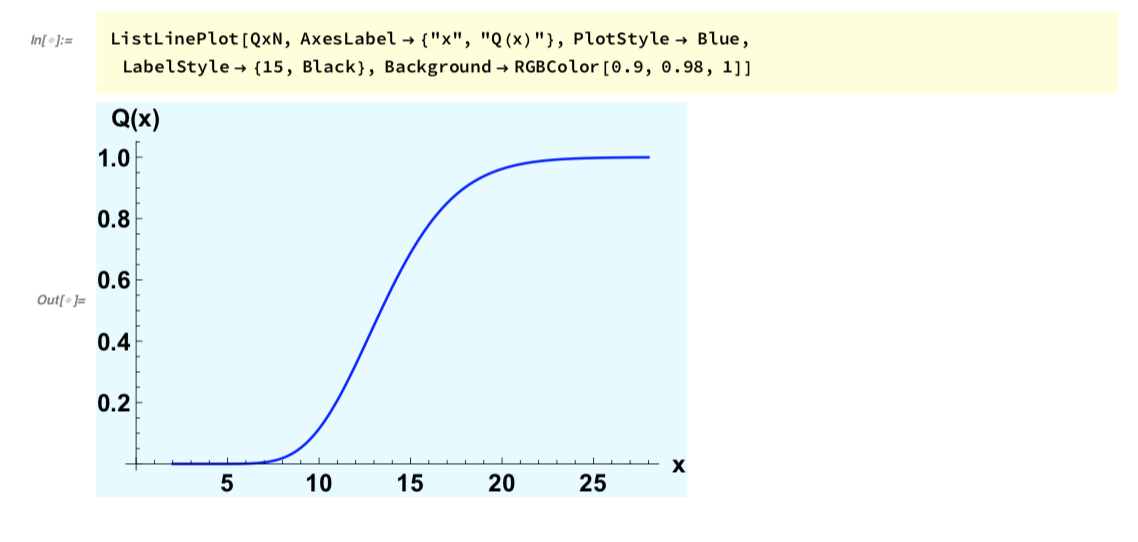}
\end{figure*}

\section*{References}
\begin{thebibliography}{99}

\bibitem{Forrester2010} Forrester P J 2010  {\it Log-Gases and Random Matrices (LMS-34)} (Princeton, NJ: Princeton University Press).

\bibitem{Mehta2004} Mehta M L 2004 {\it Random Matrices} (New York: Academic Press).

\bibitem{Handbook2011}Akemann G, Baik J and Di Francesco P (ed) 2011 {\it The Oxford Handbook of Random Matrix Theory} (Oxford: Oxford University Press)

\bibitem{Constantine1963} Constantine A G 1963 Some non-central distribution problems in multivariate analysis {\it Ann. Math. Stat.} {\bf 34} 1270

\bibitem{Anderson1963} Anderson T W 1963 Asymptotic theory for principal component analysis {\it Ann. Math. Stat.} {\bf 34} 122

\bibitem{Khatri1964} Khatri C G 1964 Distribution of the largest or the smallest characteristic root under null hypothesis concerning complex multivariate normal populations {\it Ann. Math. Stat.} {\bf 35} 1807

\bibitem{Muirhead1982} Muirhead R J 1982 {\it Aspects of Multivariate Statistical Theory} (New York: Wiley)

\bibitem{Forrester1993} Forrester P J 1993 The spectrum edge of random matrix ensembles  {\it Nucl. Phys. B} {\bf 402} 709

\bibitem{Forrester1994} Forrester P J 1994 Exact results and universal asymptotics in the Laguerre random matrix ensemble {\it J. Math. Phys.} {\bf 35} 2539

\bibitem{FH1994} Forrester P  J and Hughes T D 1994 Complex Wishart matrices and conductance in mesoscopic systems: exact results {\it J. Math. Phys.} {\bf 35} 6736

\bibitem{NF1998} Nagao T and Forrester P J 1998 The smallest eigenvalue distribution at the spectrum edge of random matrices {\it Nucl. Phys. B.} {\bf 509} 561 

\bibitem{Johansson2000} Johansson K 2000 Shape fluctuations and random matrices {\it Commun. Math. Phys.} {\bf 209} 437

\bibitem{Johnstone2001} Johnstone I M 2001  On the distribution of the largest eigenvalue in principal components analysis {\it Ann. Statist.} {\bf 29} 295

\bibitem{Forrester2007} Forrester P J 2007 Eigenvalue distributions for some correlated complex sample covariance matrices {\it J.Phys. A: Math. Theor.} {\bf 40} 11093

\bibitem{WG2013} Wirtz T and Guhr T 2013 Distribution of the smallest eigenvalue in the correlated Wishart model {\it Phys. Lett. Rev.} {\bf 111} 094101

\bibitem{MBL2008} Majumdar S N, Bohigas O and Lakshminarayan A 2008 Exact minimum eigenvalue distribution of an entangled random pure state {\it J. Stat. Phys.} {\bf 131} 33

\bibitem{AV2011} Akemann G and Vivo P 2011 Compact smallest eigenvalue expressions in Wishart-Laguerre ensembles with or without a fixed trace {\it J. Stat. Mech.: Theory Exp.} {\bf 2011} P05020

\bibitem{Majumdar2011} Majumdar S N 2010 Extreme eigenvalues of Wishart matrices: application to entangled bipartite system {\it The
Oxford Handbook of Random Matrix Theory} (Oxford: Oxford University Press) (arXiv:1005.4515)

\bibitem{KSA2017} Kumar S, Sambasivam B and Anand S 2017 Smallest eigenvalue density for regular or fixed-trace complex Wishart-Laguerre ensemble and entanglement in coupled kicked tops {\it J. Phys. A: Math. Theor.} {\bf 50} 345201  

\bibitem{Vivo2011} Vivo P 2011 Largest Schmidt eigenvalue of random pure states and conductance distribution in chaotic cavities {\it J. Stat. Mech.: Theory Exp.} {\bf 2011} P01022

\bibitem{KA2003} Kang M and Alouini M-S 2003 Largest eigenvalue of complex Wishart matrices and performance analysis of MIMO MRC systems {\it IEEE J. Sel. Areas Commun.} {\bf 21} 418

\bibitem{DMJ2003} Dighe P A, Mallik R K and Jamuar S S 2003 Analysis of transmit-receive diversity in Rayleigh fading {\it IEEE Trans. Commun.} {\bf 51} 694

\bibitem{MA2005} Maaref A and A\"{\i}ssa S 2005 Closed-form expressions for the outage and ergodic Shannon capacity of MIMO MRC systems {\it IEEE Trans. Commun.} {\bf 53} 1092

\bibitem{WTDM2012} Wei L, Tirkkonen O, Dharmawansa P and McKay M 2012  On the exact distribution of the scaled largest eigenvalue {\it Proc. IEEE Int. Conf. Commun. (ICC2012)}, pp. 2422--2426

\bibitem{PL2008} Park C S and Lee K B 2008 Statistical multimode transmit antenna selection for limited feedback MIMO systems {\it IEEE Tran. Wireless Commun.} {\bf 7} 4432

\bibitem{ZCW2009} Zanella A, Chiani M, and Win M Z 2009 On the marginal distribution of the eigenvalues of Wishart matrices {\it IEEE Trans. Commun.} {\bf 57} 1050

\bibitem{WT2011} Wei L and Tirkkonen O 2011 Analysis of scaled largest eigenvalue based detection for spectrum sensing {\it Proc. IEEE Int. Conf. Commun. (ICC2011)}, pp. 1--5

\bibitem{KSRZ2012} Kortun A, Sellathurai M, Ratnarajah T and Zhong C 2012 Distribution of the ratio of the largest eigenvalue to the trace of complex Wishart matrices {\it IEEE Trans. Signal Process.} {\bf 60} 5527

\bibitem{JHCBC2017} Jones S R, Howard S D, Clarkson I V L, Bialkowski K S and Cochran D 2017 Computing the largest eigenvalue distribution for complex Wishart matrices {\it Proc. IEEE International Conference on Acoustics, Speech and Signal Processing (ICASSP)}, pp. 3439--3443

\bibitem{TW1993} Tracy C A and Widom H 1993 Level-spacing distributions and the Airy kernel {\it Phys. Lett. B} {\bf 305} 115

\bibitem{TW1994} Tracy C A and Widom H 1994 Level-spacing distributions and the Airy kernel {\it Commun. Math. Phys.} {\bf 159} 151

\bibitem{TW1994a} Tracy C A and Widom H 1994 Level spacing distributions and the Bessel kernel {\it Commun. Math. Phys.} {\bf 161} 289

\bibitem{FS2010} Feldheim O N and Sodin S. 2010 A universality result for the smallest eigenvalues of certain sample covariance matrices {\it Geom. Funct. Anal.} {\bf 20} 88

\bibitem{VMB2007} Vivo P, Majumdar S N and Bohigas O 2007 Large deviations of the maximum eigenvalue in Wishart random matrices {\it J. Phys. A: Math. Theor.} {\bf 40} 4317

\bibitem{MV2009} Majumdar S N and Vergassola M 2009 Large deviations of the maximum eigenvalue for Wishart and gaussian random matrices {\it Phys. Rev. Lett.} {\bf 102} 060601

\bibitem{KC2010} Katzav E and Castillo I P 2010  Large deviations of the smallest eigenvalue of the Wishart-Laguerre ensemble {\it Phys. Rev. E} {\bf 82}(R) 040104

\bibitem{Forrester2012} Forrester P J 2012 Large deviation eigenvalue density for the soft edge Laguerre and Jacobi $\beta$-ensembles {\it J. Phys. A: Math. Theor.} {\bf 45} 145201

\bibitem{Edelman1989} Edelman A 1989 Eigenvalues and condition numbers of random matrices {\it Ph.D. thesis} MIT

\bibitem{Edelman1991} Edelman A 1991 The distribution and moments of the smallest eigenvalue of a random matrix of Wishart type {\it Lin. Alg. Appl.} {\bf 159} 55

\bibitem{Forrester1993a} Forrester P J 1993 Recurrence equations for the computation of correlations in the $1/r^2$ quantum many-body system {\it J. Stat. Phys.} {\bf 72} 39

\bibitem{Forrester2010a} Forrester P J and Ito M 2010 Difference system for Selberg correlation integrals {\it J. Phys. A: Math. Theor.} {\bf 43} 175202

\bibitem{FR2012} Forrester P J and Rains E M 2012 A Fuchsian matrix differential equation for Selberg correlation integrals {\it Commun. Math. Phys.} {\bf 309} 771

\bibitem{FT2019} Forrester P J and Trinh A K 2019 Optimal soft edge scaling variables for the Gaussian and Laguerre even $\beta$ ensembles {\it Nucl. Phys. B} {\bf 938} 621

\bibitem{Kumar2019} Kumar S 2019 Recursion for the smallest eigenvalue density of $\beta$-Wishart-Laguerre ensemble {\it J. Stat. Phys.} {\bf 175} 126

\bibitem{Beenakker1997} Beenakker C W J 1997 Random-matrix theory of quantum transport {\it Rev. Mod. Phys.} {\bf 69} 731

\bibitem{BM1994} Baranger H U and Mello P A 1994 Mesoscopic transport through chaotic cavities: a random S-matrix theory approach {\it Phys. Rev. Lett.} {\bf 73} 142

\bibitem{JPB1994} Jalabert R A, Pichard J-L and Beenakker C W J 1994 Universal quantum signatures of chaos in ballistic transport {\it Europhys. Lett.} {\bf 27} 255

\bibitem{MK2004} Mello P A and Kumar N 2004 {\it Quantum Transport in Mesoscopic Systems} (Oxford: Oxford University Press). 

\bibitem{SWS2007} Sommers H-J, Wieczorek W and Savin D V 2007 Statistics of conductance and shot-noise power for chaotic cavities {\it Acta Phys. Pol. A} {\bf 112} 691

\bibitem{KSS2009} Khoruzhenko B A, Savin D V and Sommers H-J 2009 Systematic approach to statistics of conductance and shot-noise in chaotic cavities {\it Phys. Rev. B} {\bf 80} 125301

\bibitem{VMB2008} Vivo P, Majumdar S N and Bohigas O 2008 Distributions of conductance and shot noise and associated phase transitions {\it Phys. Rev. Lett.} {\bf 101} 216809

\bibitem{KP2010} Kumar S and Pandey A 2010 Conductance distributions in chaotic mesoscopic cavities {\it J. Phys. A: Math. Theor.} {\bf 43} 285101

\bibitem{JG1972} Johnson D E and Graybill F A 1972 An analysis of a two-way model with interaction and no replication {\it J. Amer. Stat. Assoc.} {\bf 67} 862

\bibitem{Davis1972} Davis A W 1972 On the ratios of the individual latent roots to the trace of a Wishart matrix {\it J. Multivar. Anal.} {\bf 2} 440

\bibitem{SKC1973} Schuurmann F J, Krishnaiah P R and Chattopadhyay A K 1973 On the distributions of the ratios of the extreme roots to the trace of the Wishart matrix {\it J. Multivar. Anal.} {\bf 3} 445

\bibitem{KS1974} Krishnaiah P R and Schuurmann F J 1974 On the evaluation of some distributions that arise in simultaneous tests for the equality of the latent roots of the covariance matrix {\it J. Multivar. Anal.} {\bf 4} 265

\bibitem{BS2006} Besson O and Scharf L L 2006 CFAR matched direction detector {\it IEEE Trans. Signal Process.} {\bf 54} 2840

\bibitem{BDMN2011} Bianchi P, Debbah M, Maida M and Najim J 2011 Performance of statistical tests for single-source detection using random matrix theory {\it IEEE Trans. Inf. Theory} {\bf 57} 2400

\bibitem{Nadler2011} Nadler B 2011 On the distribution of the ratio of the largest eigenvalue to the trace of a Wishart matrix {\it J. Multivar. Anal.} {\bf 102} 363

\bibitem{ZS2001} \.{Z}yczkowski K and Sommers H-J 2001 Induced measures in the space of mixed quantum states {\it J. Phys. A: Math. Gen} {\bf 34} 7111

\bibitem{James1975} James A T, {\it Special functions of matrix and single argument in statistics}, in Theory and Applications of Special Functions, edited by R. A. Askey (Academic, New York, 1975): pp. 497-520

\bibitem{Mathematica} Wolfram Research Inc. Mathematica Version 11.3 (2018)

\bibitem{Imry1986} Imry Y 1986 Active transmission channels and universal conductance fluctuations {\it Europhys. Lett.} {\bf 1} 249

\bibitem{AS1986} Altshuler B L and Shklovski\u{\i} B I 1986 Repulsion of energy levels and conductivity of small metal samples {\it Sov. Phys. JETP} {\bf 64} 127

\bibitem{SMMP1991} Stone A D, Mello P A, Muttalib K A and Pichard J- L, in Mesoscopic Phenomena in Solids, edited by B. L. Altshuler, P. A. Lee, and R. A. Webb (North-Holland, Amsterdam, 1991): p. 369

\bibitem{Beenakker2015} Beenakker C W J 2015 Random-matrix theory of Majorana fermions and topological superconductors {\it Rev. Mod. Phys.} {\bf 87} 1037

\bibitem{Forrester2006} Forrester P J 2006  Quantum conductance problems and the Jacobi ensemble {\it J. Phys. A: Math. Gen.} {\bf 39} 6861

\bibitem{Brouwer1995} Brouwer P W 1995 Generalized circular ensemble of scattering matrices for a chaotic cavity with nonideal leads {\it Phys. Rev. B} {\bf 51} 16878

\bibitem{VK2012} Vidal P and Kanzieper E 2012 Statistics of reflection eigenvalues in chaotic cavities with nonideal leads 2012 {\it Phys. Rev. Lett.} {\bf 108} 206806

\bibitem{Landauer1957} Landauer R 1957 Spatial variation of currents and fields due to localized scatterers in metallic conduction {\it IBM J. Res. Dev.} {\bf 1} 223

\bibitem{FL1981} Fisher D S and Lee P A 1981 Relation between conductivity and transmission matrix {\it Phys. Rev. B} {\bf 23}(R) 6851

\bibitem{KP2010a} Kumar S and Pandey A 2010 Jacobi crossover ensembles of random matrices and statistics of transmission eigenvalues {\it J. Phys. A: Math. Theor.} {\bf 43} 085001

\end{thebibliography}
\end{document}